\documentclass[letterpaper, 12 pt, conference]{ieeeconf}  

\IEEEoverridecommandlockouts                              
\usepackage{prg-stuff}

\title{\LARGE \bf
A Taxonomy of Privacy Constructs \\for Privacy-Sensitive Robotics
}

\author{Matthew Rueben$^{1}$, Cindy M. Grimm$^{1}$, Frank J. Bernieri$^{2}$, and William D. Smart$^{1}$
\thanks{$^{1}$Authors Rueben, Grimm, and Smart are with the Robotics Group at Oregon State University.\newline
        {\tt\small ruebenm@oregonstate.edu}\newline
        {\tt\small cindy.grimm@oregonstate.edu}\newline
        {\tt\small bill.smart@oregonstate.edu}\newline
        $^{2}$Author Bernieri is with the School of Psychological Science at Oregon State University.\newline
        {\tt\small frank.bernieri@oregonstate.edu}}%
}

\begin{document}

\maketitle
\thispagestyle{empty}
\pagestyle{empty}

\begin{abstract}

The introduction of robots into our society will also introduce new
concerns about personal privacy. In order to study these concerns, we
must do human-subject experiments that involve measuring
privacy-relevant constructs. This paper presents a taxonomy of privacy
constructs based on a review of the privacy literature. Future work in
operationalizing privacy constructs for HRI studies is also discussed.

\end{abstract}

\section{Introduction}
In the future, robots promise to become pervasive in our society, if
not ubiquitous. Already the advent of the Internet, webcams, and
affordable mobile devices has changed the way we think about personal
privacy; robots, many of which can move around the world, will further
change this paradigm. Whereas webcams are tethered to stationary
computers and mobile devices are carried by people, robots can go
places and collect data without direct human aid, and perhaps even
unbeknownst to humans altogether. This poses a new threat to personal
privacy in all its senses: control over private information, the right
not to be recorded, personal space and solitude, and so on.

We call the study of privacy issues in robotics and how to mitigate
them, ``privacy-sensitive robotics.'' This area of research itself
probably belongs in the field of human-robot interaction, or HRI. In
order to study privacy-sensitive robotics, we must do human-subject
experiments; tackling a human-robot interaction problem without
consulting the humans is a doomed endeavor. The problem is,
``privacy'' has many meanings, so testing hypotheses about ``privacy''
is impossible without being much more specific and choosing just a
small part of ``privacy'' to work with. This paper presents a
breakdown of ``privacy'' into many constructs (i.e., abstract ideas)
organized into a hierarchical taxonomy based on a review of the
privacy literature.

\section{Background: Robots and Privacy}
Privacy-sensitive robotics can be thought of as a subset of human-robot interaction. Goodrich and Schultz have surveyed human-robot interaction \cite{goodrich_human-robot_2007} and Fong has surveyed socially-interactive robots \cite{fong_survey_2003}. The focus in both of these surveys is on autonomous robot behaviors, although in some cases, autonomy is shared between the human and the robot. Why do autonomous robots pose a privacy concern for humans? Research has revealed that humans often interact socially with machines. This phenomenon is often stated as \emph{``Computers Are Social Actors'' (CASA)} \cite{nass_computers_1994}. Any robot, then, can function as a social actor during a human-robot interaction. Broad discussions of privacy issues that are specific to robotics are only recently beginning to be published, especially outside of the robotics discipline. Calo gives a good overview as well as some newer insights \cite{calo_robots_2010}. 

Privacy is important in all human cultures \cite{altman_privacy_1977}, although different cultures have different norms for privacy and different mechanisms for enforcing those norms. 

Unfortunately, people are not always rational when they make decisions about privacy \cite{acquisti_privacy_2005}. Researchers have even had to measure privacy \emph{attitudes} separately from privacy \emph{behaviors} because of how poorly people put their privacy preferences into action \cite{berendt_privacy_2005}. In research, the value of privacy is often quantified in monetary terms, and has been shown to depend on the context (i.e., whether privacy protection is being increased or decreased \cite{acquisti_what_2013}). 

If robots can function as social actors in whichever human culture they inhabit, we want to study how we can enculturate robots with respect to our privacy norms. We call research that studies these questions ``privacy-sensitive robotics.''

\section{A Taxonomy of Privacy Constructs for Human-Robot Interactions}
This section lays out our taxonomy of privacy constructs and summarizes the key literature behind it. Definitions of terms are to be found via the references where not defined hereafter. The taxonomy is as follows:
\begin{enumerate}

\item Privacy (see Leino-Kilpi et al. \cite{leino-kilpi_privacy:_2001} for subdivision)
  \begin{enumerate}

  \item Informational (see Solove \cite{solove_understanding_2008} for subdivision)
    \begin{enumerate}
    \item Invasion
    \item Collection
    \item Processing
    \item Dissemination
    \end{enumerate}

  \item Physical
      \begin{enumerate}
      \item Personal Space \cite{westin_privacy_1967}
      \item Territoriality \cite{westin_privacy_1967,newell_perspectives_1995,burgoon_communication_1982} (see Altman \cite{altman_environment_1975} for subdivision)
        \begin{enumerate}
        \item Intrusion
        \item Obtrusion
        \item Contamination
        \end{enumerate}
      \item Modesty \cite{allen_privacy_2011}
      \end{enumerate}

  \item Psychological
    \begin{enumerate}
    \item Interrogation \cite{westin_privacy_1967}
    \item Psychological Distance \cite{hall_hidden_1966}
    \end{enumerate}
    
  \item Social 
    \begin{enumerate}
    \item Association \cite{allen_privacy_2011}
    \item Crowding/Isolation \cite{altman_environment_1975}
    \item Public Gaze \cite{austin_privacy_2003}
    \item Solitude \cite{allen_privacy_2011} (see Westin \cite{westin_privacy_1967} for subdivision)
    \item Intimacy
    \item Anonymity
    \item Reserve
    \end{enumerate}

  \end{enumerate}
\end{enumerate}

\subsection{The Literature behind the Taxonomy}
We recommend the Stanford Encyclopedia of Philosophy article on privacy by Judith DeCew as a comprehensive guide to the definition of privacy \cite{decew_privacy_2013}, especially in law and philosophy. Most of the references in this section we owe to the bibliography from that article. 

{\bf 1.a-d} Leino-Kilpi et al. \cite{leino-kilpi_privacy:_2001} divide privacy as follows:

\begin{enumerate}
\item Physical privacy, over personal space or territory
\item Psychological privacy, over thoughts and values
\item Social privacy, over interactions with others and influence from them
\item Informational privacy, over personal information
\end{enumerate}

\begin{figure}
  \centering
  \includegraphics[width=0.45\textwidth]{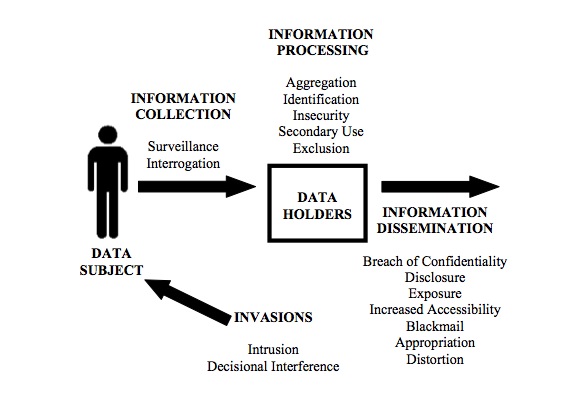}
  \caption{Daniel Solove's visual ``model'' of his taxonomy of (informational) privacy \cite{solove_understanding_2008}. }
  \label{fig:solove-taxonomy}
\end{figure}

{\bf 1.a.i-iv} Informational privacy refers to privacy concerns about personal information. In 1960, William Prosser divided (informational) privacy into four parts. His formulation continues to be referenced today. Briefly, Prosser divides (informational) privacy into intrusion, public disclosure, false light, and appropriation. These mean the following. First, intrusion into one's private affairs includes trespassing, search, and remote intrusion such as wire tapping. Second is public disclosure of private facts. Third is publicly portraying the victim in a false light, e.g., by misattributing to the victim a statement or opinion. Fourth is appropriation, or pretending to be the victim for one's own advantage. Daniel Solove has constructed a taxonomy of privacy concepts based on Prosser's formulation. It is shown in Figure~\ref{fig:solove-taxonomy} as a general overview of informational privacy concerns. We use the highest level of Solove's hierarchy for 1.a.i-iv. 

{\bf 1.b.i-ii} Privacy could be defined in terms of one's personal space or territory. These concepts are found readily in proxemics literature as well as in psychology and ethology (i.e., animal behavior studies) in general, but are not often connected with privacy. Patricia Newell includes territoriality in her review of \textit{Perspectives on Privacy} \cite{newell_perspectives_1995}, although she also cites a study that separates between the two \cite{edney_distinguishing_1976}. Leino-Kilpi et al. \cite{leino-kilpi_privacy:_2001} define physical privacy as being over personal space and territory, and Westin also mentions it when he links human privacy ideas with animal behavior \cite{westin_privacy_1967}. 

Social psychologist Irwin Altman pulls together the related concepts of privacy, personal space, territoriality, and crowding \cite{altman_environment_1975}. His book, along with Burgoon's article \cite{burgoon_communication_1982} (discussed below), is a good foundation for \emph{environmental and spatial} factors related to privacy. 

Judee Burgoon presents a communication perspective on privacy, including territoriality, in a broad survey \cite{burgoon_communication_1982}. She argues that more ``physical'' privacy could consist of blocking more communication channels, including sight, sound, and even smell (e.g., the smell of food being cooked next door). We would add further channels enabled by technology: phone calls, text messages, Facebook posts, and the like. Alternatively, Burgoon writes that to have more territory, higher-quality territory (e.g., better-insulated walls), and more unquestioned control over that territory is to enjoy more physical privacy. 

{\bf 1.b.iii} Allen lists \emph{modesty} as an important physical privacy concern in medical settings, especially from the philosophical standpoints of Christian ethics and virtue ethics \cite{allen_privacy_2011}. Modesty may drive patients to request same-sex or even same-race doctors. 

{\bf 1.c.i} According to Westin's account of privacy in U.S. law, the right to privacy swelled in the late 1900's \cite{westin_privacy_1967}. The Supreme Court continued to try cases in which new technologies created privacy concerns beyond physical entry and tangible items. According to Westin, new protections included ``associational privacy'' over group memberships (this is distinct from 1.d.i) and ``political privacy'' over unfair questioning on account of political positions. 

{\bf 1.c.ii} \emph{Proxemics} can include psychological distance as well as physical distance (see Hall \cite{hall_hidden_1966} cited by Mumm and Mutlu \cite{mumm_human-robot_2011}).

{\bf 1.d.i and iv} Privacy might also include solitude, i.e., being physically removed from other people. Solitude is more than a freedom from trespassing; one needn't be at home to desire solitude. Anita Allen includes solitude in her article on privacy and medicine \cite{allen_privacy_2011}. In the medical setting, the sick often want to be comforted by company, but also to have some time alone. This could be especially true for patients with terminal illnesses, who might want to reflect on their lives and make some important decisions. In such cases we tend to respect their wishes. Allen also mentions ``associational privacy,'' the ability to choose one's own company \cite{allen_privacy_2011}. She notes that patients do not merely desire intimacy, but rather ``selective intimacy'' with certain loved ones, and this is an aspect of privacy to consider. 

{\bf 1.d.ii} Altman calls both crowding and isolation failures to regulate the amount of interaction with others \cite{altman_environment_1975}. It may seem odd to call social isolation a privacy issue, but it is a logical conclusion from within Altman's theory of privacy (see Appendix).

{\bf 1.d.iii} Lisa Austin offers a more nuanced definition of privacy: freedom from ``public gaze'' \cite{austin_privacy_2003}. She argues that this updated definition deals with the problem of new technologies to which older definitions of privacy do not object. In particular, Austin is concerned about cases wherein people know they are under surveillance, about the collection of non-intimate but personal information (e.g., in data mining), and about the collection of personal information in public. She claims that other, older definitions of privacy do not agree with our intuition that these technologies (could) invade our privacy by denying us our freedom from ``public gaze.''

{\bf 1.d.iv-vii} Alan Westin lists four different states of privacy: solitude, anonymity, intimacy (i.e., being alone \emph{with} someone), and reserve (i.e., keeping to oneself) \cite{westin_privacy_1967}. 

\section{Future Work}
This taxonomy takes the broad concept of privacy and breaks it into more specific constructs. We have split the single trunk into what we see as its main branches, and some of those branches have also been shown to fork off, too. To study privacy in human-robot interaction (e.g., in human-subject experiments), we need the leaves of this privacy tree. Unlike the trunk and branches, the leaves are no longer abstract constructs; instead, they are concrete measures. For example, one operationalization of personal information collection (1.a.ii) would be whether someone knows your social security number -- a simple, binary measure. Other measures might be contextual, e.g., given that you are alone in a room with a PR2 robot staring at you, do you feel comfortable changing your shirt? This comfort level, a proxy for modesty (1.b.iii), could be measured, for example, by a questionnaire. All such measures would tap the extent to which a person's privacy has been preserved or violated. 


\bibliographystyle{IEEEtran}
\bibliography{privacy-survey}

\begin{thebibliography}{10}
\providecommand{\url}[1]{#1}
\csname url@rmstyle\endcsname
\providecommand{\newblock}{\relax}
\providecommand{\bibinfo}[2]{#2}
\providecommand\BIBentrySTDinterwordspacing{\spaceskip=0pt\relax}
\providecommand\BIBentryALTinterwordstretchfactor{4}
\providecommand\BIBentryALTinterwordspacing{\spaceskip=\fontdimen2\font plus
\BIBentryALTinterwordstretchfactor\fontdimen3\font minus
  \fontdimen4\font\relax}
\providecommand\BIBforeignlanguage[2]{{%
\expandafter\ifx\csname l@#1\endcsname\relax
\typeout{** WARNING: IEEEtran.bst: No hyphenation pattern has been}%
\typeout{** loaded for the language `#1'. Using the pattern for}%
\typeout{** the default language instead.}%
\else
\language=\csname l@#1\endcsname
\fi
#2}}

\bibitem{goodrich_human-robot_2007}
M.~A. Goodrich and A.~C. Schultz, ``Human-robot interaction: a survey,''
  \emph{Foundations and trends in human-computer interaction}, vol.~1, no.~3,
  pp. 203--275, 2007.

\bibitem{fong_survey_2003}
T.~Fong, I.~Nourbakhsh, and K.~Dautenhahn, ``A survey of socially interactive
  robots,'' \emph{Robotics and autonomous systems}, vol.~42, no.~3, pp.
  143--166, 2003.

\bibitem{nass_computers_1994}
C.~Nass, J.~Steuer, and E.~R. Tauber, ``Computers {Are} {Social} {Actors},'' in
  \emph{Proceedings of the {SIGCHI} {Conference} on {Human} {Factors} in
  {Computing} {Systems}}, ser. {CHI} '94.\hskip 1em plus 0.5em minus
  0.4em\relax New York, NY, USA: ACM, 1994, pp. 72--78.

\bibitem{calo_robots_2010}
R.~Calo, ``Robots and privacy,'' \emph{ROBOT ETHICS: THE ETHICAL AND SOCIAL
  IMPLICATIONS OF ROBOTICS, Patrick Lin, George Bekey, and Keith Abney, eds.,
  Cambridge: MIT Press, Forthcoming}, 2010.

\bibitem{altman_privacy_1977}
I.~Altman, ``Privacy {Regulation}: {Culturally} {Universal} or {Culturally}
  {Specific}?'' \emph{Journal of Social Issues}, vol.~33, no.~3, pp. 66--84,
  1977.

\bibitem{acquisti_privacy_2005}
A.~Acquisti and J.~Grossklags, ``Privacy and rationality in individual decision
  making,'' \emph{IEEE Security \& Privacy}, no.~1, pp. 26--33, 2005.

\bibitem{berendt_privacy_2005}
B.~Berendt, O.~Günther, and S.~Spiekermann, ``Privacy in e-commerce: stated
  preferences vs. actual behavior,'' \emph{Communications of the ACM}, vol.~48,
  no.~4, pp. 101--106, 2005.

\bibitem{acquisti_what_2013}
A.~Acquisti, L.~K. John, and G.~Loewenstein, ``What is privacy worth?''
  \emph{The Journal of Legal Studies}, vol.~42, no.~2, pp. 249--274, 2013.

\bibitem{leino-kilpi_privacy:_2001}
H.~Leino-Kilpi, M.~Valimaki, T.~Dassen, M.~Gasull, C.~Lemonidou, A.~Scott, and
  M.~Arndt, ``Privacy: {A} {Review} of the {Literature},'' \emph{International
  Journal of Nursing Studies}, vol.~38, pp. 663--671, 2001.

\bibitem{solove_understanding_2008}
D.~J. Solove, ``Understanding privacy,'' 2008.

\bibitem{westin_privacy_1967}
A.~F. Westin, \emph{Privacy and {Freedom}}.\hskip 1em plus 0.5em minus
  0.4em\relax New York, NY: Athenaeum, 1967.

\bibitem{newell_perspectives_1995}
P.~B. Newell, ``Perspectives on {Privacy},'' \emph{Journal of Environmental
  Psychology}, vol.~15, pp. 87--104, 1995.

\bibitem{burgoon_communication_1982}
J.~Burgoon, ``Privacy and communication,'' in \emph{Communication Yearbook 6},
  M.~Burgoon, Ed.\hskip 1em plus 0.5em minus 0.4em\relax Routledge, 1982,
  no.~6.

\bibitem{altman_environment_1975}
I.~Altman, \emph{The Environment and Social Behavior: Privacy, Personal Space,
  Territory, and Crowding.}\hskip 1em plus 0.5em minus 0.4em\relax Monterey,
  CA: Brooks/Cole Publishing Company, 1975.

\bibitem{allen_privacy_2011}
A.~Allen, ``Privacy and {Medicine},'' in \emph{The {Stanford} {Encyclopedia} of
  {Philosophy}}, spring 2011~ed., E.~N. Zalta, Ed., 2011.

\bibitem{hall_hidden_1966}
E.~T. Hall, \emph{The {Hidden} {Dimension}}.\hskip 1em plus 0.5em minus
  0.4em\relax Doubleday, Garden City, 1966.

\bibitem{austin_privacy_2003}
L.~Austin, ``Privacy and the {Question} of {Technology},'' \emph{Law and
  Philosophy}, vol.~22, no.~2, pp. 119--166, 2003.

\bibitem{decew_privacy_2013}
J.~DeCew, ``Privacy,'' in \emph{The {Stanford} {Encyclopedia} of {Philosophy}},
  fall 2013~ed., E.~N. Zalta, Ed., 2013.

\bibitem{edney_distinguishing_1976}
J.~J. Edney and M.~A. Buda, ``Distinguishing territoriality and privacy: {Two}
  studies,'' \emph{Human Ecology}, vol.~4, no.~4, pp. 283--296, 1976.

\bibitem{mumm_human-robot_2011}
J.~Mumm and B.~Mutlu, ``Human-robot proxemics: physical and psychological
  distancing in human-robot interaction,'' in \emph{Proceedings of the 6th
  international conference on {Human}-robot interaction}.\hskip 1em plus 0.5em
  minus 0.4em\relax ACM, 2011, pp. 331--338.

\bibitem{nissenbaum_privacy_2004}
H.~Nissenbaum, ``Privacy as contextual integrity,'' \emph{Wash. L. Rev.},
  vol.~79, p. 119, 2004.

\bibitem{moore_privacy:_2003}
A.~D. Moore, ``Privacy: its meaning and value,'' \emph{American Philosophical
  Quarterly}, pp. 215--227, 2003.

\bibitem{inness_privacy_1992}
J.~C. Inness, \emph{Privacy, intimacy, and isolation}.\hskip 1em plus 0.5em
  minus 0.4em\relax Oxford University Press, 1992.

\end{thebibliography}

\begin{appendix}
Here we add some very important theories about privacy that didn't make it into this paper because they are too general, but are essential for understanding privacy as a whole (and hence any one construct in our taxonomy). 

Altman's theory defines privacy as a \emph{boundary regulation process} wherein people try to achieve their ideal privacy state by using certain \emph{mechanisms} to regulate interaction with others \cite{altman_environment_1975}. Notice how this definition allows privacy to sometimes mean \emph{more} interaction with others, and sometimes \emph{less} interaction; successfully switching between the two is the key. Along these lines, Altman calls privacy a \emph{dialectic process}, i.e., a contest between two opposing forces -- withdrawal and engagement -- which alternate in dominance. Hence, privacy to Altman is \emph{dynamic} in that the desired level of engagement changes over time for a given individual. This theory is necessary for understanding Altman's discussion of personal space, territoriality, and crowding.

Helen Nissenbaum's approach to privacy, which she calls ``contextual integrity,'' focuses on the idea that different norms of information gathering and dissemination are observed in different contexts \cite{nissenbaum_privacy_2004}. Privacy is violated in a given context when the norms for information gathering or dissemination within that context are broken. Nissenbaum argues that some scenarios, especially public surveillance, are intuitively felt by many to be potential privacy violations, and that while U.S. legal policy overlooked these scenarios (at time of writing), ``contextual integrity'' does a better job of accounting for our intuitive concerns \cite{nissenbaum_privacy_2004}. 

Adam Moore defines privacy as, ``control over access to oneself and information about oneself'' \cite{moore_privacy:_2003}. This is a ``control-based'' definition of privacy, in which it doesn't matter whether somebody accesses you or your information, but rather whether you can control that access. Control-based definitions account for situations in which someone invites others into his close company, or willingly gives out personal information. These actions would violate privacy if privacy is the state of being let alone, or of having all your personal information kept to yourself. But authors holding to control-based definitions of privacy maintain that the person in question is still in control, so there's no violation; this especially makes sense in the legal context.

Julie Inness wrote the book on privacy as it relates to intimacy \cite{inness_privacy_1992}. She proposes that intimate interactions must be \emph{motivated} by liking, love, or care in order to be intimate. As evidence she points to Supreme Court decisions wherein constitutional privacy protection was conferred to issues of the family and sexual health due to the personal, emotional impacts that made those issues intimate. In this way, Inness seems to define privacy as the protection of intimate matters, where intimacy comes from the motivation and not the behavior itself (e.g., kissing is not automatically intimate). She recognizes that this definition of intimacy is subjective, making legal rulings more difficult.

\end{appendix}

\end{document}